\documentclass[preprint]{elsarticle}
\usepackage{graphicx}
\usepackage{dcolumn}
\usepackage{bm}
\usepackage{natbib}
\usepackage{caption}
\usepackage{subcaption}
\usepackage{rotating}

\newcommand{\comments}[1]{}

\newcommand{\subscript}[1]{\ensuremath{_{\textrm{#1}}}}

\begin{document}
\title{First Principles Modeling of the Temperature Dependent Ternary Phase Diagram for the Cu-Pd-S System}
\author[CMU]{William Paul Huhn\corref{cor1}}
\ead{wph@andrew.cmu.edu}
\author[CMU]{Michael Widom}
\ead{widom@andrew.cmu.edu}
\author[NETL,URS]{Michael C. Gao}
\cortext[cor1]{Corresponding author}
\address[CMU]{Department of Physics, Carnegie Mellon University, Pittsburgh, PA 15213, USA}
\address[NETL]{National Energy Technology Laboratory, Albany, OR 97321, USA}
\address[URS]{URS Corporation, Albany, OR 97321, USA}

\begin{abstract}
As an aid to the development of hydrogen separation membranes, we predict the temperature 
dependent phase diagrams using first principles calculations combined with thermodynamic 
principles.  Our method models the phase diagram without empirical fitting parameters.  By 
applying thermodynamic principles and solid solution models, temperature-dependent features 
of the Cu-Pd-S system can be explained, specifically solubility ranges for substitutions in 
select crystalline phases.  Electronic densities of states calculations explain the relative 
favorability of certain chemical substitutions.  In addition, we calculate sulfidization 
thresholds for the Pd-S\subscript{2} system and activities for the Cu-Pd binary in temperature 
regimes where the phase diagram contains multiple solid phases.
\end{abstract}
\maketitle
\section{Introduction}
Sulfidization impedes the use of palladium-based membranes for hydrogen sequester and capture.  
Due to the presence of hydrogen sulfide (H\subscript{2}S) in any practical feed gas, sulfur 
rapidly covers the surface of the membrane, rendering its catalytic properties inert.  Pure 
palladium membranes have limited application for hydrogen sequester and capture due to 
irreversible formation of bulk sulfides~\cite{CHChen10} and embrittlement due to lattice 
distortion caused by hydrogen atoms~\cite{HAmandusson01} diffusing through the bulk.  
Alloying palladium with other metals, such as silver~\cite{SUemiya91} or gold~\cite{CHChen10}, 
fixes many of the problems associated with palladium membranes, but these materials still suffer 
from sulfidization and are expensive.  In the 2000s it was discovered that alloying palladium 
with copper~\cite{Iyoha07,OBrien10} dramatically reduces the rate of sulfidization and further 
decreases the price of these membranes, though still not eliminating sulfidization.   In order 
to realize hydrogen capture and sequester on an industrial scale, new membranes with different 
chemical compositions that resist sulfidization need to be devised, and their reactivity to 
sulfur needs to be examined.  Computational methods offer an attractive way to rapidly prototype 
different chemical families.  

In this paper we computationally model the temperature-dependent phase diagram for the 
Cu-Pd-S system using first principles methods.  Prior experimental work~\cite{MollerMakovicky99} 
determined the ternary phase diagram for the Cu-Pd-S system.  We seek a simple and 
semi-quantitative model to gain insight into the essential features of the system, in 
contrast to complicated models with many fitting parameters that can yield close 
agreement with experiment at the cost of obscuring the underlying physics. 

First, we describe our method for calculating ternary phase diagrams and apply it to Cu-Pd-S.  
Our model begins with first principles total energies, moves to focus on substitutional entropy, then examines vibrations.  Comparison with 
experimental results shows general agreement.  Certain details of energetics are explored through
examination of the electronic density of states.  We then apply our model to predict sulfidization
thresholds for Pd interacting with S dimers and activities in Cu-Pd binaries.

\section{Methods and Calculations}
In order to calculate a phase diagram using first principles methods, all plausible crystal structures must
be considered, in principle.  In the case of Cu-Pd-S, a wealth of 
experimental crystallographic data gives an excellent starting point, especially for the binary compounds.  
We generalize these binaries to ternary structures formed from substitutions of the ternary species into either lattice sites or 
interstitial sites.  Once the T=0K total energies for all likely crystalline structures have been calculated, it is 
straightforward to generate a T=0K phase diagram by creating the convex hull of total energies.

Subramanian and Laughlin~\cite{CuPdPhase91} determined the phase diagram for Cu-Pd experimentally, 
and Li \emph{et al.} created a CALPHAD model~\cite{Li08}.  Above T=871K, 
the entire composition range of solid Cu-Pd exists in fcc solid solution.  Below this temperature  
other solid phases are present.  $\beta$-CuPd has a CsCl structure and exhibits a Cu-rich composition 
range extending to a composition of 58\% Cu.  Trimarchi and Zunger~\cite{Trimarchi08} showed using an 
evolutionary algorithm that first principles calculations predict cP2 to be the expected ground 
state at equiatomic composition.  At copper concentrations near 75\%, phases based on 1D and 
2D long period superstructures (LPS) compete.  Studies using ab initio methods to examine this 
region of the phase diagram have already been performed~\cite{Barthlein09,Ceder90}.  As we are 
primarily concerned with the palladium-rich side of the phase diagram, we only calculate the 
ideal 1D LPS, Cu\subscript{3}Pd.tP28, (using a notation of [chemical formula].{Pearson symbol}) 
and ignore the 2D LPS and phase solubility ranges in the 1D LPS.

Determination of the phase diagram for Cu-S~\cite{CuSPhaseDiagram07} using first principles 
calculations is problematic, as it contains an abundance of phases, many of which are 
non-stoichiometric or have large unit cells.  We include only Cu-S structures with well-defined 
crystallographic refinements, as we do not expect this discrepancy to affect our results on the 
palladium-rich side of the phase diagram.  The Pd-S phase diagram~\cite{PdSPhaseDiagram92} is 
relatively simple, containing Pd\subscript{16}S\subscript{7}, Pd\subscript{4}S, PdS, and 
PdS\subscript{2} phases as line compounds at low temperature, with Pd\subscript{3}S stabilized 
at T=829K.  We summarize the phase diagrams here only for comparison; our method does not require 
pre-existing knowledge of the phase diagrams.

In order to add temperature dependence to the phase diagrams, we introduce free energy models for 
select phases of interest, all of which are based on substitutional solid solution models with 
sublattice filling (where for some structures, the "sublattice" is the entire lattice).  When 
calculating the ternary phase diagram and the activities, we work in the Gibbs ensemble, where 
we specify temperature, pressure, and chemical concentration.  We neglect pressure dependence in 
the solid phases' free energies, as solid compressibilities are low.  Phonon free energies are 
included, in the harmonic approximation, giving free energies of form 
\begin{equation}
G_{tot} = G_{config} + G_{phonon}.
\label{eq:Gtot}
\end{equation} 
Electronic free energy differences between competing phases of interest (Pd\subscript{16}S\subscript{7} 
versus Pd\subscript{4}S and well-ordered fcc-CuPd versus $\beta$-CuPd at 50\% Cu), were found to be 
less than 1.5 meV/atom at temperatures as high as 1000 K.  These differences are within the margin of
error expected for our fits for analytic free energy models, and to simplify our modeling we have 
chosen to neglect $G_{elect}$.

For lattice models, two types of defects are common, substitutions and interstitials.  Due to
atomic size differences, the only interstitials expected to be relevant are sulfur interstitials
in Cu-Pd phases.  We calculate interstitial defect costs using the well-ordered structures Cu.cF4, 
Pd.cF4, Cu\subscript{50}Pd\subscript{50}.cF4, and CuPd.cP2, with sulfur occupying a single interstitial
site for various supercell sizes.  The least unfavorable substitution, octahedral in 
Cu\subscript{50}Pd\subscript{50}.cF4, lies approximately 2 eV above the convex hull with significant 
distortion, and thus we expect contributions to sulfur solubility by occupation of interstitials to be 
negligible.

We here summarize the configurational free energy G\subscript{config} of a simple substitutional 
solid solution model, valid in the dilute limit.  In a substitutional solid solution, we have a 
set of site classes i, each with N\subscript{i} sites, an associated enthalpic cost of performing 
a chemical substitution of species $\alpha$ into site class i, $\Delta$E\subscript{i,$\alpha$}, 
and a substitution concentration of x\subscript{i,$\alpha$}.  In the substitutional solid solution 
model, it is assumed that each substitution is uncorrelated with all other chemical substitutions, 
and all spatial configurations have identical energies.  This implies a linear dependence of 
$\Delta H$ with respect to composition, giving an enthalpic contribution of
\begin{equation}
\Delta H=\Delta H_0+\sum_{i,\alpha}x_{i,\alpha}N_{i}\Delta E_{i,\alpha},
\label{eq:HConfFull}
\end{equation}
where $\Delta H_0$ denotes the enthalpy of the base structure with no substitutions.  Including an ideal entropy of mixing
\begin{equation}
\Delta S=-k_{B}\sum_{i,\alpha}N_{i}[x_{i,\alpha}\log(x_{i,\alpha})+(1-x_{i,\alpha})\log(1-x_{i,\alpha})],
\label{eq:SConfFull}
\end{equation}
leads to a Gibbs free energy $\Delta G=\Delta H-T*\Delta S$.
Here, $\Delta$H\subscript{0} and $\Delta$E\subscript{i,$\alpha$} can be determined from first-principles calculations. 
However, we control only the total impurity concentration for a species $\alpha$, x\subscript{$\alpha$}, not the number of substitutions 
in a given site class x\subscript{i,$\alpha$}.  Since we work with a fixed number of sites, we fix the total concentration of substitutions, and
thus minimize the free energy over all x\subscript{i,$\alpha$} subject to the constraint of species number conservation.

To better elucidate the model, we restrict to the special case where 
only one type of substitution (with an enthalpic cost of $\Delta E$) 
is allowed in only one sublattice containing $N_{0}$ sites.  
A case with two sublattices will be considered later.  Other special cases
include the binary regular solution model ($N=N_{0}$, where $N$ is the total
number of atoms) and the well-known binary ideal solution model ($N=N_{0}$, 
$\Delta E = 0$).  The configurational free energy takes the form 
\begin{equation}
\Delta G(N',T)=\Delta H_0 + \Delta E*N' +kTN_{0}(x'\log x' + (1-x')\log(1-x'))
\label{eq:GConfSimple}
\end{equation}
where $N'$ is the number of substitutions performed and $x'=N'/N_{0}$ is the concentration of substitutions in the sublattice.  
The appropriate quantity to consider for thermodynamics is the free energy per atom, $\Delta g=\Delta G/N$, expressed in terms 
of the $x=N'/N$, which can be shown to be 
\begin{equation}
\Delta g(x,T)=\Delta h_0 + \Delta E*x - T\Delta S
\label{eq:GConfPerAtomSimple}
\end{equation}  
where $\Delta h_0 = \Delta H_{0}/N$, and $\gamma = N_{0}/N$ is a measure of the number density of sublattice sites.  Of particular
importance is the entropic contribution to the free energy,
\begin{equation}
-T\Delta S=k_{B}T\gamma(x\log x + (\gamma-x)\log(\gamma-x) - \gamma\log\gamma)
\label{eq:SConfPerAtomSimple}
\end{equation}  
which we use for all phases modeled throughout this paper.

In order for a given substitution type to appreciably affect the free energy, the two necessary features
are large $\gamma$ values, corresponding to a sublattice dense in the total lattice and thus a significant entropic contribution
from the sublattice, and small positive (or any negative) enthalpic cost $\Delta h_{0}$ for the substitutions, such that at a given
temperature the entropic portion of the free energy can overcome the enthalpic cost (or reinforce the enthalpic gain).

We use VASP (the Vienna Ab-Initio Simulation Package) 
~\cite{KresseHafner93,KresseFurthmuller96} a plane wave ab-initio package 
implementing PAW potentials~\cite{Blochl94} to determine total energies.  Previous 
work by Hu et al.~\cite{HuGaoDoganKingWidom10} compared the accuracy of different 
pseudopotentials in the Pd-S system, and recommended PBEsol~\cite{PerdewEtAl08} 
as most suitable for calculations involving sulfur.   The total energy calculations 
were performed by fully relaxing atomic positions and lattice parameters 
until energies converged to within 0.1 meV/atom.  Subsequent convergence in 
k-point density was performed.  A common energy cutoff of 273 eV 
was used.  In order to model solubility ranges, calculations were performed 
both in unit cells and larger supercells.  All phonon calculations were done 
using density functional perturbation theory, and were performed in at least 
a 2x2x2 supercell of the unit cell, with the exception of 
Pd\subscript{16}S\subscript{7}.cI46, where due to its large size only a 2x2x2 
supercell of the primitive cell was used.  Though a supercell method with 
chemical ordering is being used to model phonon free energies for certain disordered
phases, Wu et al~\cite{Wu03} found that the high-temperature vibrational entropy 
difference between L1$_{2}$ and SQS8 configurations is only 0.03$k_{B}$, within 
their margin of error and corresponding to an upper bound on the free energy 
difference of 2.6 meV/atom at 1000K.  Table \ref{tab:structures} shows a list 
of structures and their total energies.

\begin{center}
	\begin{table}
	\begin{tabular}{ |c|c|c|c|c|c|c| }
		\hline
		Phase & Pearson Symbol & Prototype & Group \# & Space Group & $\Delta$H & dE  \\
		\hline
		Cu~\cite{Owen33} & cF4 & Cu & 225 & Fm$\overline{3}$m & 0 & 0 \\
		\hline
		Pd~\cite{Harris72} & cF4 & Cu & 225 & Fm$\overline{3}$m & 0 & 0 \\
		\hline
		S\subscript{2} Dimer & N/A & N/A & N/A & N/A & 0 & 0 \\
		\hline
		Cu\subscript{3}Pd~\cite{Okamura70} & tP28 & Cu\subscript{3}Pd & 99 & P4mm & -103.0 & 0 \\
		\hline
		CuPd & cP2 & ClCs & 221 & Pm$\overline{3}$m & -116.9 & 0 \\
		\hline
		Cu\subscript{2}S~\cite{Evans79,Leon81} & mP144 & Cu2S & 14 & P2$_{1}$/c & -496.4  & 0 \\
		\hline
		CuS~\cite{FjellvagEtAl88} & oC24 & CuS & 63 & Cmcm & -715.4 & 0 \\
		\hline
		CuS\subscript{2}~\cite{KingPrewitt79} & cP12 & FeS\subscript{2} & 205 & Pa$\overline{3}$ & -821.5 & 0 \\
		\hline
		Pd\subscript{4}S~\cite{Gronvold62} & tP10 & Pd\subscript{4}Se & 114 & P$\overline{4}$2$_{1}$C & -418.9 & 0\\
		\hline
		Pd\subscript{16}S\subscript{7}~\cite{Matkovic76} & cI46 & Pd\subscript{16}S\subscript{7} & 217 & I$\overline{4}$3m & -589.7 & 0 \\ 
		\hline
		PdS~\cite{Gaskell37} & tP16 & PdS & 84 & P4$_{2}$/m & -892.8 & 0 \\
		\hline
		PdS\subscript{2}~\cite{Gronvold57} & oP12 & PdSe\subscript{2} & 61 & Pbca & -952.4 & 0 \\
		\hline
		CuS~\cite{FjellvagEtAl88} & hP12 & CuS & 120 & I$\overline{4}$c2 & -712.9 & 2.5 \\
		\hline
		Cu\subscript{2}S~\cite{Janosi64} & tP12 & Cu\subscript{2}S & 96 & P4$_{3}$2$_{1}$2 & -479.3 & 3.3 \\
		\hline
		Cu\subscript{4}Pd~\cite{GuymontGratias76} & cP4 & AuCu\subscript{3} & 221 & Pm$\overline{3}$m & -99.7 & 3.4 \\
		\hline
		Pd\subscript{3}S~\cite{Rost68} & oC16 & Pd\subscript{3}S & 40 & Ama2 & -492.8 & 8 \\
		\hline
		Cu\subscript{50}Pd\subscript{50}~\cite{Soutter71} & cF4 & Cu & 225 & Fm$\overline{3}$m & -106.9 & 10 \\
		\hline
		Cu\subscript{4}Pd~\cite{Geisler54} & tP20 & Cu4Pd & 84 & P4$_{2}$/m & -69.6 & 12.8 \\
		\hline
		Cu\subscript{2}Pd\subscript{3}S\subscript{4} & mP18 & Cu\subscript{2}Pd\subscript{3}Se\subscript{4} & 14 & P2\subscript{1}/c& -743.0 & 13.2 \\
		\hline
		Cu\subscript{2}S~\cite{Yamamoto91} & cF12 & CaF\subscript{2} & 225 & Fm$\overline{3}$m & -503.1 & 20.6 \\
		\hline
		Cu & cI2 & W & 229 & Im$\overline{3}$m & 36.3 & 36.3 \\
		\hline
	\end{tabular}
	\caption{A partial list of calculated structures, focusing on "base" structures with no substitutions.  All energetic quantities are given in units of meV/atom.  Cu\subscript{2}Pd\subscript{3}S\subscript{4} is a hypothetical structure based on Cu\subscript{2}Pd\subscript{3}Se\subscript{4}.mP18~\cite{Topa06}. Cu.cI2 is a hypothetical structure used to model the $\beta$-CuPd phase.  $\Delta$H and dE are defined in eqs. \ref{eq:EOF} and \ref{eq:dE}}
	\label{tab:structures}
	\end{table}	
\end{center}
\section{Results and Discussion}
\subsection{Phase Diagrams}
Given total energies, we calculate the enthalpy of formation for our structures at T=0K,
\begin{equation}
\Delta h=\Delta h_{0}-\sum_{i}x_{i}\Delta h_{i},
\label{eq:EOF}
\end{equation}
where $\Delta$h\subscript{0} is the calculated T=0K total energy for the base structure, 
$x_{i}$ is the composition of a given species $i$ in the structure, and 
$\Delta h_{i}$ is the T=0K total energy for pure species $i$ in a chosen reference structure.  By 
convention, this reference structure is chosen to be the equilibrium structure at T=0K, so that 
at T=0K all enthalpies of formation for the pure species vanish.  In the Cu-Pd-S system, the standard reference structures would be
Cu.cF4 for copper, Pd.cF4 for palladium, and S.mP48 for sulfur.  

As we are interested in interaction of gaseous sulfur with copper-palladium membranes, we have chosen to use 
gaseous sulfur dimers rather than the standard S.mP48 when calculating
phase diagrams.  In the case of the ternary phase diagram, all structures of interest have relatively low sulfur 
composition, so the mismatch in reference phase compared to experimental phase diagrams is negligible.  

The quantity of importance for constructing temperature dependent phase diagrams is 
\begin{equation}
dE=\Delta h-[\Delta h],
\label{eq:dE}
\end{equation}
where [$\Delta$h] is the enthalpy of formation of the convex hull 
at a structure's stoichiometry.  By definition, this quantity is non-negative,
and it vanishes only when the structure lies on the convex hull.  This 
quantity is a measure of how far above the convex hull a particular 
structure lies at T=0K.  Structures with large dE's require large 
entropic contributions to free energy in order to overcome enthalpic costs.  

\begin{figure}
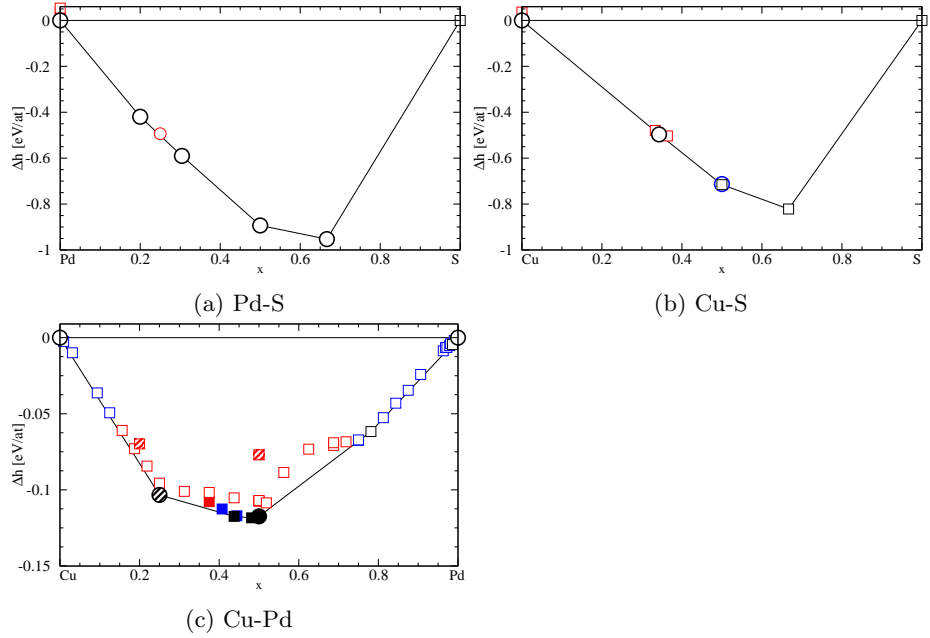

\begin{subfigure}[b]{0.5\textwidth}
	\includegraphics[trim = 0mm 0mm 0mm 0mm, clip, width=\textwidth]{Pd-S.eps}
	\caption{Pd-S}
\end{subfigure}
\begin{subfigure}[b]{0.5\textwidth}
	\includegraphics[trim = 0mm 0mm 0mm 0mm, clip, width=\textwidth]{Cu-S.eps}
	\caption{Cu-S}
\end{subfigure}
\begin{subfigure}[b]{0.5\textwidth}
	\includegraphics[trim = 0mm 0mm 0mm 0mm, clip, width=\textwidth]{Cu-Pd.eps}
	\caption{Cu-Pd}
\end{subfigure}
\caption{(color online) Binary enthalpies of 
formation for Pd-S, Cu-S, and Cu-Pd.  The x axis corresponds to concentration, 
and y axis corresponds to enthalpy of formation. Black symbols denote stable 
structures, blue correspond to structures witin 3 meV of the convex hull, and red to 
structures above 3 meV.  Circles with thick borders are known low temperature stable
structures, circles with thin borders are known high temperature stable structures, and squares denotes all other 
cases (including hypothetical structures).  For Cu-Pd, to better distinguish phases,
filled symbols correspond to bcc structures, empty symbols to fcc, and dashed for the 
three other structures (Cu\subscript{3}Pd.tP28, Cu\subscript{4}Pd.tP20, and the SQS configuration 
for fcc at 50\% Pd).}
\label{fig:binaryphase}
\end{figure}

Figure \ref{fig:binaryphase} shows the T=0K binary enthalpies of formation.  
Observe that palladium and sulfur react strongly, as can be seen 
by the maximum enthalpy of formation of 0.95 eV/atom.  This strong enthalpy of 
formation reflects the tendency of pure palladium membranes to sulfidize.  Next, copper and sulfur react 
almost as strongly, with a maximum enthalpy of formation of 0.82 eV/atom.  Finally, 
copper and palladium have a maximum enthalpy of formation of only 0.12 eV/atom, indicating that the interspecies
binding between copper and palladium is relatively weak.  This is
expected from the Cu-Pd phase diagram, as all known phases are variants of fcc (the T=0, p=0 structure for both Cu and 
Pd) or bcc (which is reached from fcc by a martensic transition~\cite{BrunoGinatempoGiuliano01}).  
We find that the known structure CuS.oC24 is mechanically unstable, with an imaginary phonon mode that stabilizes a slight monoclinic 
distortion, and this distorted monoclinic structure, CuS.mC24, is the true ground state structure according to DFT.
The stable phases predicted correspond to known low temperature stable phases, with the exception of 
CuS$_{2}$.tP12, which is not observed stable at temperatures as low as 300K (its stability is mostly 
likely due to the use of S\subscript{2} as a reference structure), and CuS.mC24, the experimentally observed stable structure being CuS.hP12.    

\begin{figure}
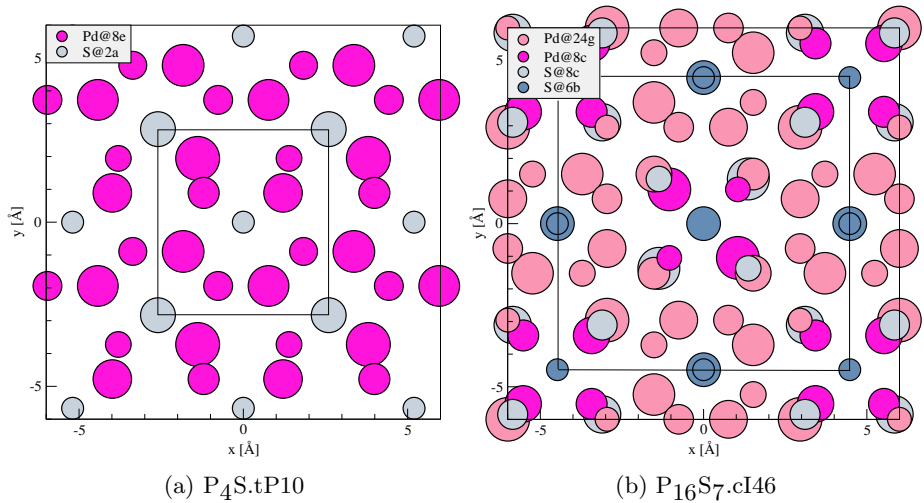

\begin{subfigure}[b]{0.5\textwidth}
	\includegraphics[trim = 0mm 0mm 0mm 10mm, clip, width=\textwidth]{Pd4StP10.eps}
	\caption{P\subscript{4}S.tP10}
\end{subfigure}
\begin{subfigure}[b]{0.5\textwidth}
	\includegraphics[trim = 0mm 0mm 0mm 10mm, clip, width=\textwidth]{Pd16S7cI46.eps}
	\caption{P\subscript{16}S\subscript{7}.cI46}
\end{subfigure}
\caption{(color online) Crystal structures for P\subscript{4}S.tP10 and P\subscript{16}S\subscript{7}.cI46.  The unit cells are outlined.  Atomic sizes indicate vertical height.}
\label{fig:pd16s7-pd4s}
\end{figure}

Experimental results~\cite{MollerMakovicky99} show Pd\subscript{16}S\subscript{7}.cI46 
and Pd\subscript{4}S.tP10 having solubility ranges for copper, which implies the existence of substitutional 
or interstitial defects.  Pd\subscript{16}S\subscript{7}.cI46 contains four 
Wyckoff site classes:  8c and 24g sites containing palladium and 8c and 6b sites
containing sulfur (see Fig. \ref{fig:pd16s7-pd4s}).  For Pd\subscript{4}S.tP10 there are only two site classes, an 8g containing palladium and 2a 
containing sulfur. 

\begin{center}
	\begin{figure}
	\includegraphics[width=0.75\textwidth]{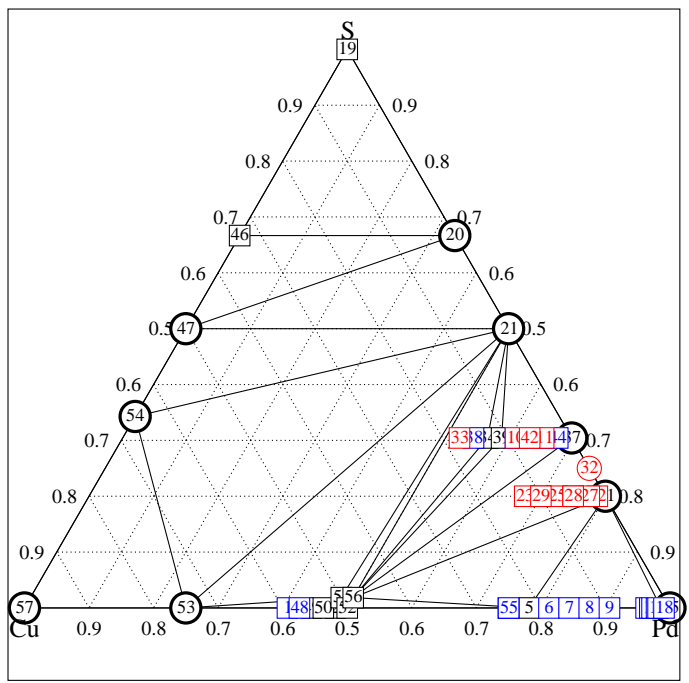}
	\caption{(color online) The calculated T=0K phase diagram for the Cu-Pd-S system.  Plotting symbols as in Fig. \ref{fig:binaryphase}.}
	\label{fig:ternary0K}
	\end{figure}
\end{center}

Shown in Figure \ref{fig:ternary0K} is our calculated T=0K phase diagram 
for the Cu-Pd-S system (supporting data is in Table \ref{tab:subs}).  We find the 
8c site class in Pd\subscript{16}S\subscript{7} is favorable to copper substitution.  At T=0K it is enthalpically 
favorable to occupy 6 of the 8 possible lattice sites, with a sudden upturn for further filling of the 8c site class.  
The only other substitutions that are found favorable are S and Cu in $\beta$-CuPd, Cu in Pd.cF4, and Pd in CuS\subscript{2}.  
Substitutions of Cu in the $\beta$-CuPd phase have a solubility range from 50\% Cu to 55.5\% at T=0K.  We find the $\beta$-CuPd
phase is only favorable to Cu substitutions on the Pd sublattice, and not Pd substitutions on the Cu sublattice, which is
experimentally supported by observations that the palladium content of $\beta$-CuPd never exceeds 50\%.
As none of the Cu- and Pd- rich structures are in competition with CuS\subscript{2}.tP12, and the entropic contribution 
is small, we ignore substitutions in CuS\subscript{2}.tP12.  We also find a small 
sulfur solubility range in the Cu sublattice of $\beta$-CuPd at T=0K, which is 
supported by experimental phase diagrams at T=673K and T=823K. 
 
While these are the only stable T=0K substitutions, we expect that at higher temperatures,
substitutions into other sites that had suitably low enthalpic cost will be stabilized by entropy:  PdS.tP16 (2c), 
Pd\subscript{16}S\subscript{7}.cI46 (24g), and Pd\subscript{4}S.tP10 (8e) sites being possible candidates (see Table \ref{tab:subs}).  
However, we chose to exclude modeling the PdS.tP16 2c substitution range, as due to the small concentration of 2c sites in 
PdS.tP16, combined with competition with other phases with strong entropic effects, we expect the concentration range to 
be negligible.    

\begin{figure}
\includegraphics[width=\textwidth]{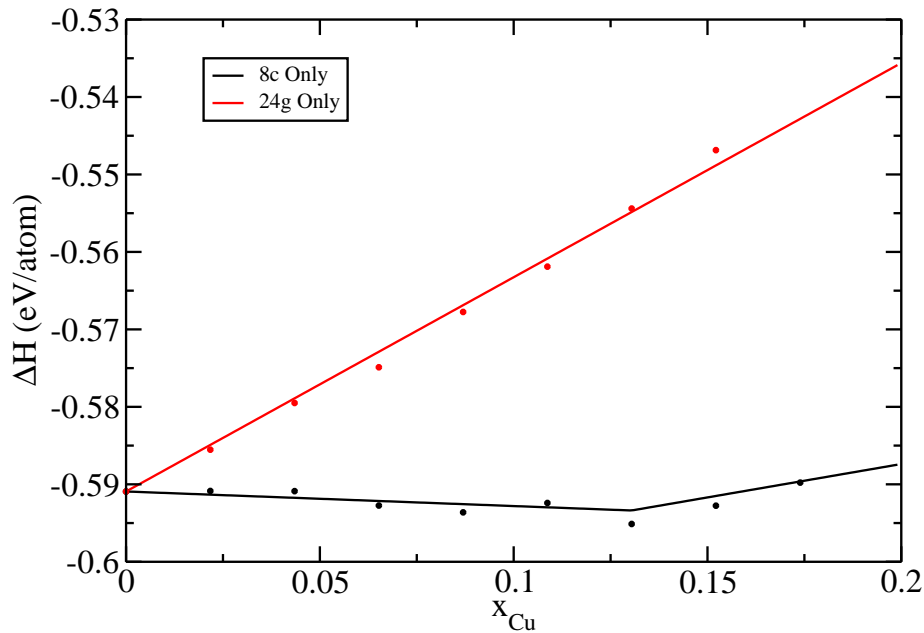}
\caption{(color online) Comparison for the enthalpy of formations for filling 8c sites versus filling 24g sites in Pd\subscript{16}S\subscript{7}.cI46.}
\label{fig:8cversus24g}
\end{figure}

To determine solubility ranges more precisely, we model phases analytically (see Table \ref{tab:phaseeqs}), 
using first principles calculation data to suggest suitable enthalpic models.  All models are based off the
regular solution model, though with modifications to better reflect the behavior of the phase.  We choose 
the models to smooth out fluctuations in the first principles calculated energies while still keeping 
the essential trends seen intact.  The phases modeled are Pd\subscript{4}S, Pd\subscript{16}S\subscript{7}, 
$\beta$-CuPd, and the CuPd-FCC phase.  In all cases we use the ideal entropy (Eq. \ref{eq:SConfPerAtomSimple}) 
for sublattice filling.  Our model for Pd\subscript{16}S\subscript{7} differs from a regular solution model in 
that we use a piecewise linear function (see Fig. \ref{fig:8cversus24g}) for the enthalpic part of the free 
energy.  The enthalpic contribution to the free energy of Pd\subscript{16}S\subscript{7} is increasingly 
favorable up to 6 substitutions (x=0.13) on the 8c sublattice, but from 6 to 8 substitutions (x=0.174) the 
favorability is reduced.  Previous crystallographic work by Matkovic~\cite{Matkovic76} observed a crystalline 
phase of Pd\subscript{16}S\subscript{7} with 14\% Cu concentration, the Cu occupying 8c sites.  However this 
work was performed at 853K and did not suggest any mechanism by which these substitutions are stabilized.  For 
Pd\subscript{4}S, we use the dilute limit form of the regular solution model ($1-x \approx 1$), though it was 
necessary to add a small quadratic term to the free energy to fit our first principle results.  Both CuPd-FCC 
and $\beta$-CuPd were fit using a first-order regular solution model, with CuPd-FCC having the entire lattice 
available for substitution, allowing usage of the standard form of a regular solution model.  $\beta$-CuPd has 
only the Pd sublattice available for substitution by Cu, which may be modeled using a regular solution model by 
replacing terms involving $x$ with terms involving $x-0.5$, where $x$ denotes the Cu concentration.  We also 
compute a SQS configuration for fcc at 50 Pd\%\cite{Wolverton01} with purely random correlations until the 8th 
nearest neighbor and find it lies above the convex hull by ~40 meV/atom.

Experimental phase diagrams show a sulfur solubility range near Cu\subscript{0.5}Pd\subscript{0.5} trending in the 
palladium-rich direction.  We find dilute S substitutions onto the Cu sublattice in $\beta$-CuPd, creating a phase 
trending in the palladium-rich direction, are stable at T=0K.  Subsequent substitutions become rapidly unfavorable.  
From this, we expect limited entropic stabilization of additional substitutions of S in $\beta$-CuPd at higher 
temperatures, leading to a roughly constant solubility range.  

To include phonon free energy in solid solution phases, we calculate phonon free energies in the 
harmonic approximation at well-ordered configurations at select compositions and linearly 
interpolate between these compositions across the phase region.  To model the free energy 
of gaseous sulfur, we use empirical data taken from the NIST-JANAF tables, as explained 
later in the paper.

\begin{center}
	\begin{table}
	\begin{tabular}{ |c|c|c|c|c| }
		\hline
		Base Structure & Site Class & Chemical Species & Enthalpic Cost & Maximum Impurity Concentration \\
		& & Change & (meV/atom) & in Site Class \\
		\hline
		$\beta$-CuPd & 2a & Pd to Cu & 0 & 12.5 \\
		\hline
		$\beta$-CuPd & 2a & Cu to S & 0 & 3.8 \\
		\hline
		$\beta$-CuPd & 2a & Pd to S & 0 & 3.8 \\
		\hline
		Pd.cF4 & 4a & Cu to Pd & 0  & 21.9 \\
		\hline
		Pd\subscript{16}S\subscript{7}.cI46 & 8c & Pd to Cu & 0 & 12.5 \\
		\hline
		Pd\subscript{16}S\subscript{7}.cI46 & 24g & Pd to Cu & 12.9 & 8.33 \\
		\hline
		Pd\subscript{4}S.tP10 & 8e & Pd to Cu & 4.7 & 3.13 \\
		\hline
		CuS\subscript{2}.cP12 & 4a & Cu to Pd & 0 & 6.25 \\ 
		\hline
		PdS.tP16 & 2c & Pd to Cu & 5.3 & 50.00 \\
		\hline
	\end{tabular}
	\caption{A list of substitutions into binary structures to produce ternary structures.  
Only low enthalpic cost substitutions are shown.  Impurity
concentration denotes the concentration in the appropriate 
sublattice, not concentration in the structure as a whole.}
	\label{tab:subs}
	\end{table}	
\end{center}

\begin{center}
	\begin{sidewaystable}
	\begin{tabular}{ |c|c|c|c|c| }
	\hline
	Phase & Substitution Type & $\gamma$ & Phonon Concentrations & $h_{config}$ \\
	 & (Cu Composition Range) & & & (eV/atom) \\
	\hline
	Pd\subscript{4}S & Cu@Pd8e (0-0.8) & 8/10 & 0, 0.8 & $ -5.350 + 1.612x + 0.225x^{2} $ \\
	\hline
	Pd\subscript{16}S\subscript{7} & Cu@Pd8c (0-0.175) & 8/46 & 0, 0.175 & $ -5.237+1.445x (x < 6/46), -5.251+1.554x (x >= 6/46) $  \\
	\hline
	$\beta$-CuPd & Cu@Pd2a (0.5 - 1) & 1/2 & 0.5, 1 & $-4.8575 + 1.773(x-0.5) + (x-0.5)(1-x)(-0.206(x-0.5)-1.526(1-x))$ \\
	\hline
	CuPd-FCC & Cu@Pd4a (0-1) & 1 & 0, 0.50, 0.75, 1 & $-5.477 + 1.469x + x(1-x)(-0.529x - 0.267(1-x))$ \\
	\hline
	\end{tabular}
	\caption{Description of our phase models.  $\gamma$ denotes the number concentration of the sublattice in the total lattice and is used for scaling the entropic contribution to energy (see equation \ref{eq:SConfPerAtomSimple}). "Phonon concentrations" denotes the concentrations for which phonon free energies were calculated.  $x$ denotes the copper composition.}
	\label{tab:phaseeqs}
	\end{sidewaystable}
\end{center}

\begin{figure}
	\begin{subfigure}[b]{0.5\textwidth}
		\includegraphics[width=\textwidth]{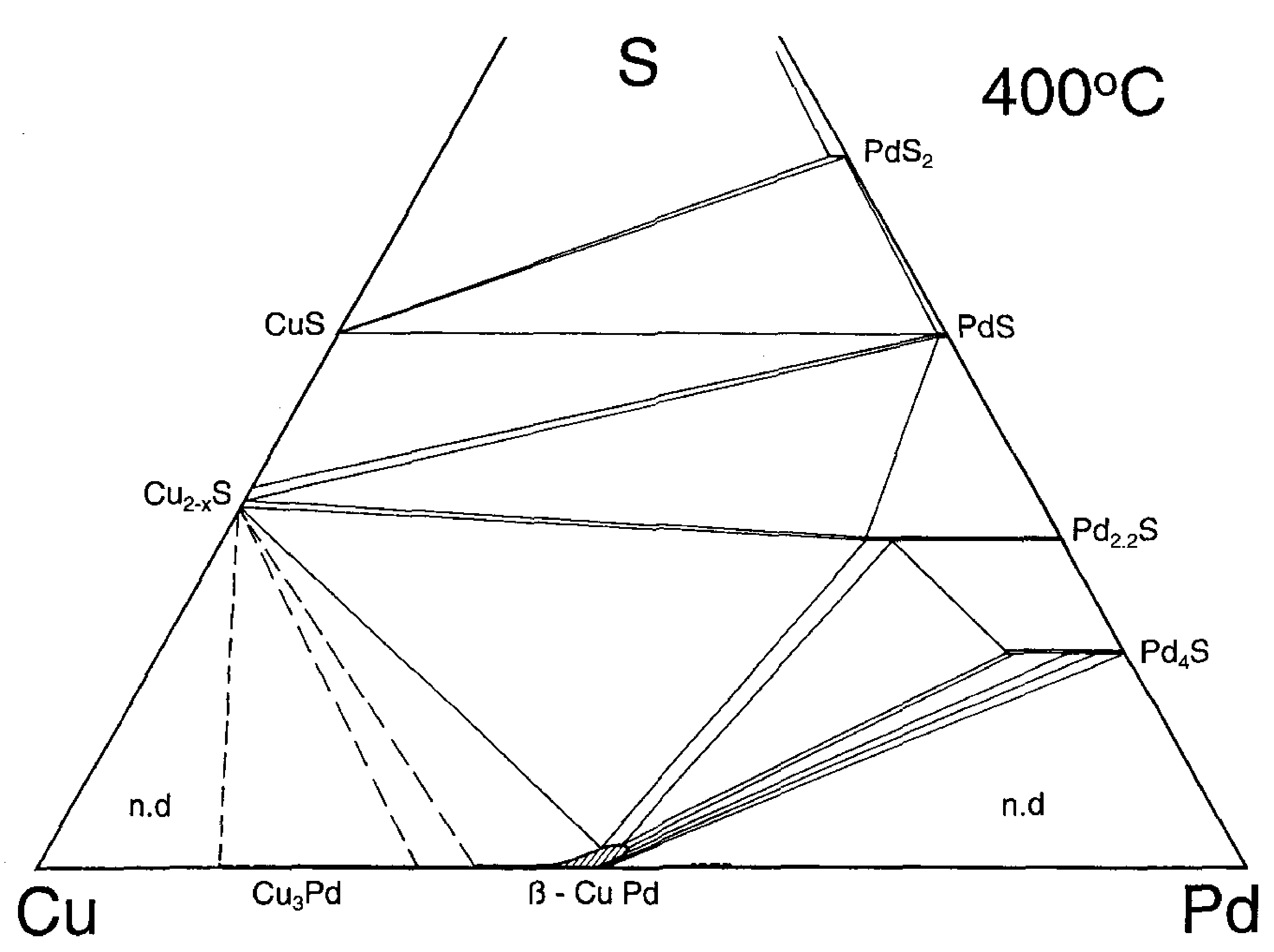}
		\caption{Experimental, T=673K}
	\end{subfigure}
	\begin{subfigure}[b]{0.5\textwidth}
		\includegraphics[width=\textwidth]{Cu-S-Pd.673K.eps}
		\caption{Calculated, T=673K}
	\end{subfigure}
	\caption{(color online) The experimental phase diagram for the Cu-Pd-S system at T=673K, and our calculated phase diagram.}
	\label{fig:ternary673K}
\end{figure}

\begin{figure}
	\begin{subfigure}[b]{0.5\textwidth}
		\includegraphics[width=\textwidth]{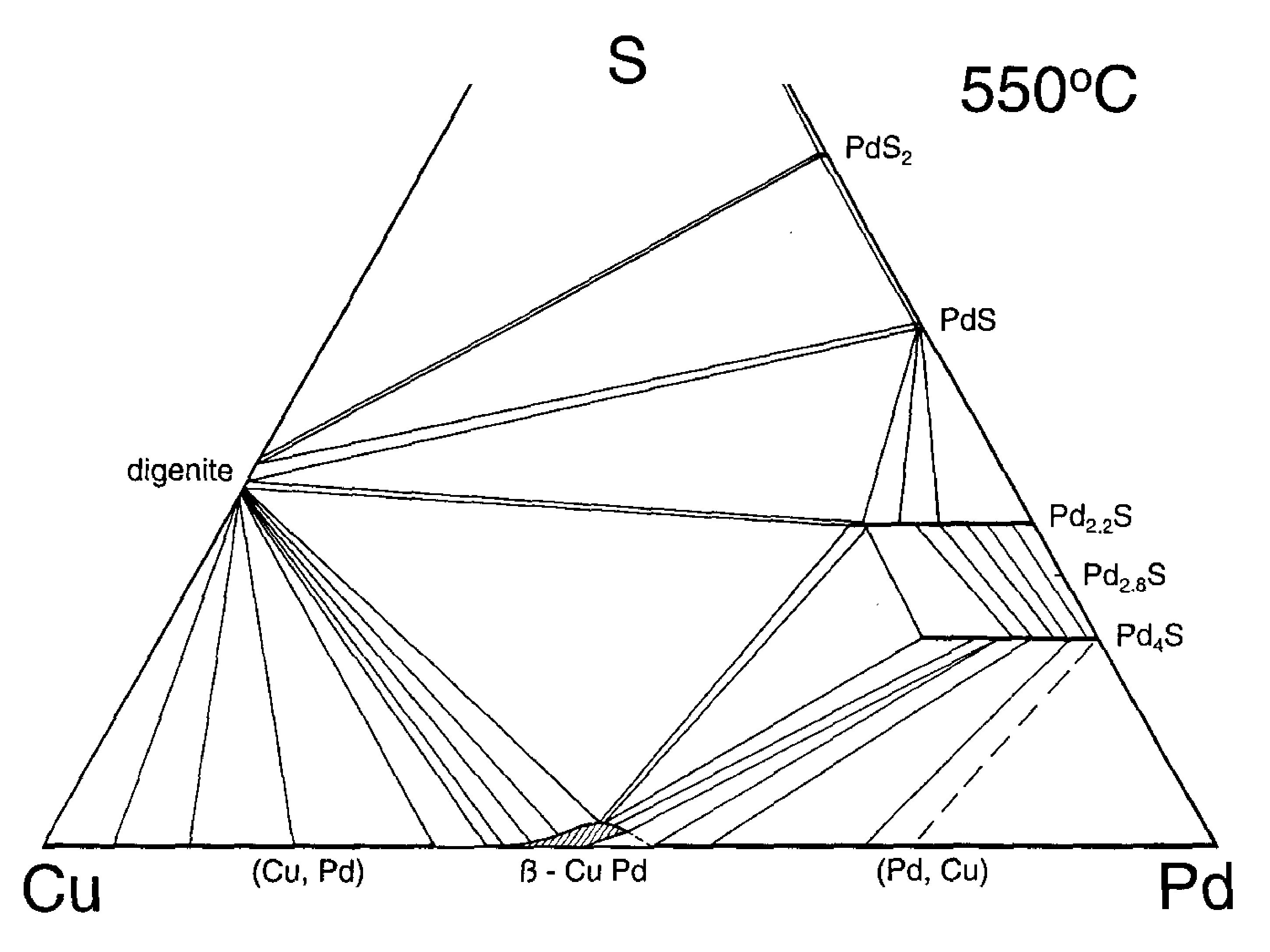}
		\caption{Experimental, T=823K}
	\end{subfigure}
	\begin{subfigure}[b]{0.5\textwidth}
		\includegraphics[width=\textwidth]{Cu-S-Pd.823K.eps}
		\caption{Calculated, T=823K}
	\end{subfigure}
	\caption{(color online) The experimental phase diagram for the Cu-Pd-S system at T=823K, and our calculated phase diagram.}
	\label{fig:ternary823K}
\end{figure}

Figures \ref{fig:ternary673K} and \ref{fig:ternary823K} compare our calculated phase diagrams 
with experiment for T=673K and 823K, respectively.  Our T=673K phase diagram matches the  
experimental results qualitatively.  Note, however, that the "Cu\subscript{2-x}S" region of the 
experimental phase diagram contains multiple structures, many of which are
non-stoichiometric, whereas ours contains only one structure.  The phase labeled 
"Cu\subscript{3}Pd" (or 1D LPS) has a slight solubility range experimentally whereas we model 
it as a well-ordered line compound.  Our T=823K diagrams show more disagreement, 
as there is only one Cu-S structure that is stable and a Pd\subscript{3}S structure is stabilized at this temperature.  More recent 
phase diagrams have Pd\subscript{3}S being stabilized at T=829K, however. This is likely due to entropic effects other than
configurational and phononic (in the harmonic approximation) that were not included.   As can be seen, Pd\subscript{3}S is 
part of our collection of calculated structures but 
is not a stable low-temperature phase.  At both temperatures we attempt to model sulfur solubility in the $\beta$-CuPd phase but find sulfur
solubility to be unstable at temperatures of interest.  This is surprising, as our T=0K phase diagram 
has a sulfur solubility range in $\beta$-CuPd similar in shape to what is experimentally observed at higher 
temperatures, and the enthalpic behavior of our sulfur substitutions supports the observed temperature independence of 
the sulfur solubility range.

The Pd\subscript{16}S\subscript{7}.cI46 copper solubility range is not 
appreciably affected by the temperature, as it is dominated by enthalpy rather than
entropy.  The 24g site is unfavorable for copper substitution.  The 
high multiplicity of the 24g site class gives a large entropic contribution 
to the free energy at high temperature relative to the 8c site class.  
However, we find that the enthalpic cost (see Figure \ref{fig:8cversus24g}) 
associated with occupying 24g sites is too high relative to the entropic contributions 
at the given temperatures and only affects the solubility range by 1\% to 2\%.  
Thus the solubility range consists almost completely of 8c sites and is inappreciably 
affected by temperature.   For this reason, our model (Table \ref{tab:phaseeqs}) only includes 8c sites and not 
24g sites.  For Pd\subscript{16}S\subscript{7}.cI46, at both measured 
temperatures the experimental copper solubility limit is 20\%, and our calculated 
value is 17.5\%, the mismatch likely due to Cu occupation of 24g sites.   
Our results explain why this solubility range appears to be independent of 
temperature;  it is not stabilized due to entropic contributions but due to enthalpy
arising from details of the electronic structure (see subsection \ref{subsec:electstruct}).

In contrast, Cu substitutions in Pd\subscript{4}S.tP10 are not enthalpically 
favorable at T=0K but are entropically stabilized at higher temperature.  As can be seen 
in the above phase diagrams, the copper solubility for Pd\subscript{4}S.tP10 is temperature-dependent 
in both the calculated and experimental work.  While our results agree qualitatively with experimental results, 
there is still a discrepancy in the copper solubility range.  For Pd\subscript{4}S.tP10, at T=673K the experimental value is 12.7\% 
and ours is 2.5\%, and for T=823K the values are 18.6\% and 5\% respectively.
Solid solution models overestimate the configurational entropy, as including correlations between substitutions 
will alter the energies of distinct configurations with the same composition and reduce the multiplicity.  We have thus 
overestimated the entropy of the Pd\subscript{16}S\subscript{7} phase, whose solubility range extends to almost complete 
filling of the 8c site, relative to the Pd\subscript{4}S phase, whose filling in the 8e sublattice is sparse.  Thus we would expect more accurate 
free energy models will lead to decreased entropy in the Pd\subscript{16}S\subscript{7} phase with comparatively small changes 
in Pd\subscript{4}S, increasing the solubility range of Pd\subscript{4}S.  This will not affect the stability of Pd\subscript{16}S\subscript{7}
as it is primarily enthalpically stabilized.  

The large solubility range in the palladium rich side of the CuPd-FCC phase 
at T=0K is also a factor in the incorrect solubility ranges for 
Pd\subscript{4}S.tP10, as we have found that altering the fitting parameters 
for the CuPd-FCC phase appreciably affects the solubility range for 
Pd\subscript{4}S.tP10.  As we also encountered an issue with sulfur solubility 
in $\beta$-CuPd, with the sulfur solubility range destabilizing relative to 
CuPd-FCC at temperature below temperatures of interest, and we have a large 
solubility range on the Pd-rich side at T=0K for the CuPd-FCC phase,  it 
appears that one major source of error is inaccurate modeling of the CuPd-FCC 
phase.  We will return to this issue in the activities portion of this paper.  

For our Cu-Pd binary at high temperatures, we find that Cu\subscript{3}Pd.tP28 
becomes unstable relative to the CuPd-FCC phase at T=827K, and $\beta$-CuPd 
becomes unstable relative to CuPd-FCC and Cu\subscript{3}Pd.tP28 at T=536K and 
a composition of 59.5\%, significantly below the experimentally observed 
temperature of T=871K but close to the experimental composition of 58\%.  This 
relative instability of $\beta$-CuPd at temperatures of interest is responsible 
for the observed instability of any sulfur solubility in $\beta$-CuPd in our 
T=673K and T=823K phase diagrams.  We will show later that first-principles
inspired corrections to the enthalpic portion of the free energy for CuPd-FCC 
will bring the phase transition closer in line with experimental results, though 
as CuPd-FCC phase has maximal entropy and $\beta$-CuPd phase (with only
Cu substitutions in Pd sites) has minimal entropy in near the equiatomic 
composition region, it is likely that the assumption of ideal entropy in the FCC 
phase also plays a role.

But even with these discrepancies, it is clear that we have predicted major details 
of the Cu-Pd-S ternary phase diagram, specifically the mechanism for stabilization 
of the solubility ranges of Pd\subscript{4}S.tP10 and Pd\subscript{16}S\subscript{7} 
as well as details about the stability of the $\beta$-CuPd system, using only first 
principles calculations with no empirical fitting.

\subsection{Electronic Structure}
\label{subsec:electstruct}
The question remains why the Pd\subscript{16}S\subscript{7}.cI46 8c site class favors copper in the first 6 
substitutions, but substitution in the final two 8c sites are unfavorable.  
Consider the constituent binaries:  copper binds less strongly with sulfur
than palladium, and copper binding with palladium is relatively weak, so we would expect copper substitution into Pd-S
 binaries to be slightly unfavorable.  If it were to turn out to be favored, then we would expect Cu to fully occupy
the 8c class.  First-principles methods provide electronic structure information that can resolve this issue.

\begin{figure}
\includegraphics[width=\textwidth]{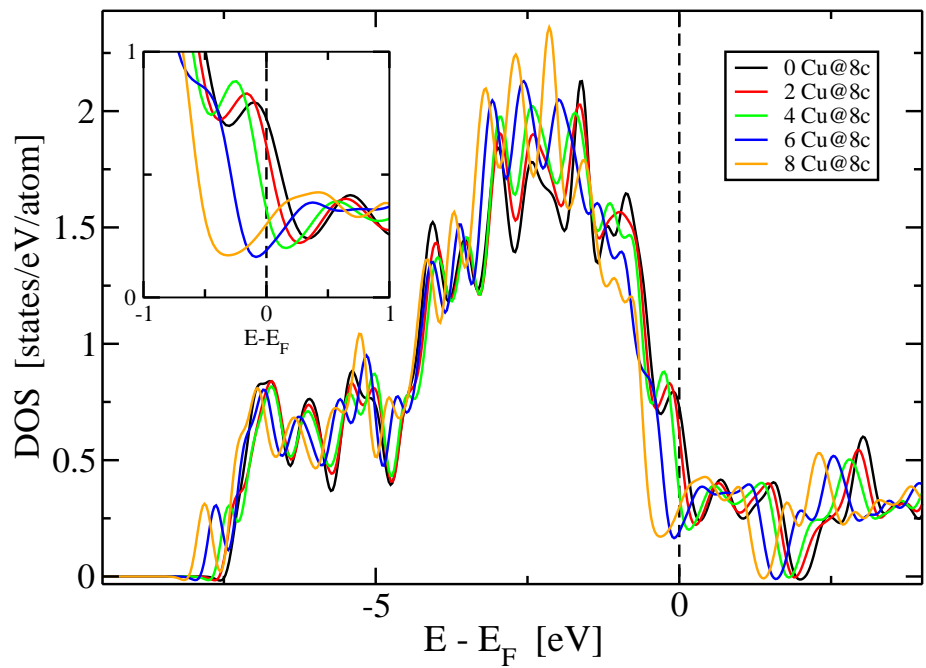}
\caption{(color online) Calculated electronic densities of states for Pd\subscript{16}S\subscript{7}.cI46 with copper substitutions at the palladium 8c sites, with the Fermi level E\subscript{F} (dashed line) set to zero.  The inset enlarges the pseudogap region.}
\label{fig:Pd16S7eDOS}
\end{figure}
Figure \ref{fig:Pd16S7eDOS} shows the electronic densities of states for substitutions in Pd\subscript{16}S\subscript{7}.cI46.  Each colored 
solid line corresponds to the electronic density of states for a different number of substitutions 
in the 8c site class of Pd\subscript{16}S\subscript{7}.cI46.  It can be seen that the densities of 
states follow a rigid band model.  Here copper is being substituted for palladium, and copper has 
one more electron in its valence shell than palladium.  Under a rigid band model, performing a 
substitution adds one electron per unit cell without appreciably changing the valence electronic 
eigenstates, increasing the Fermi level relative to the fixed band structure.
  
In all cases, a pseudogap lies near the Fermi level.  In the case of no substitutions, the pseudogap lies to 
the right of the Fermi level.  As the number of copper substitutions increases, the increased electron count 
drive the pseudogap to the left, towards the Fermi level.  At six substitutions the Fermi level lies almost 
directly at the minimum of the pseudogap.  This is a well-known stabilization mechanism for alloys~\cite{Friedel88}.
As the number of substitutions increases above six, the Fermi level is driven away from the pseudogap, 
destabilizing the structure and leading to the sudden increase in the enthalpy of formation (see Fig. 
\ref{fig:8cversus24g}).  This pseudogap stabilization mechanism at T=0K, combined with large enthalpic cost 
of 24g substitutions, explains the temperature independence of the Pd\subscript{16}S\subscript{7}
solubility range.

\subsection{Predominance Diagrams}
We now calculate the predominance diagram for the Pd-S\subscript{2} system.  We work with a fixed quantity of palladium interacting
with a reservoir of gaseous S\subscript{2}, with temperature and pressure being free parameters.  For an ideal gas,
this implies that we control the chemical potential of the gas, and thus we work in the semi-grand canonical 
ensemble:
\begin{equation}
\Sigma(T,p,{x}) = G(T,p,{x}) - \mu_{S_{2}}N_{S_{2}}
\label{eq:SemiGrand}
\end{equation}
where G is the previously defined Gibbs free energy, $\mu\subscript{S\subscript{2}}$ is the chemical potential for the sulfur dimers, and
$N_{S_{2}}=N_{S}/2$ is the half the number of sulfur atoms in a given phase (as the chemical potential is for sulfur dimers).  
To determine sulfidization thresholds, we calculate the semi-grand canonical free energy for all phases of interest, Pd.cF4, Pd\subscript{4}S.tP10,
 Pd\subscript{16}S\subscript{7}.cI46, Pd\subscript{3}S.oC16, and PdS.tP16.  The phase with the lowest free energy 
will be stable.  All Pd-containing phases of interest are line compounds, containing only total energy and phonon contributions.

\begin{figure}
\includegraphics[width=\textwidth]{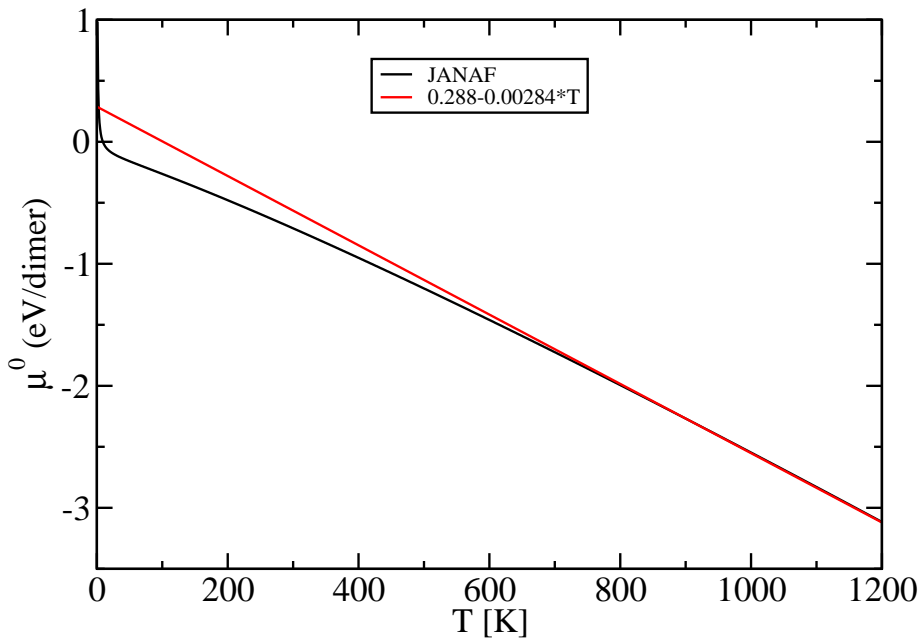}
\caption{(color online) The chemical potential of S\subscript{2} dimers.  Black is NIST-JANAF, and red is our linear fit.}
\label{fig:JANAFfit}
\end{figure}

The chemical potential for S\subscript{2} complicates the calculation, as it is a 
gaseous phase.  We incorporate empirical results into our method, in a manner 
consistent with the results obtained from first principles calculations.  For the free energy of the gas, we have
\begin{equation}
\mu(T,p)=\mu^{0}(T)+kT\log\frac{p}{p_{0}}.
\label{eq:MuSDimer}
\end{equation}
Here $\mu$ is the chemical potential per molecule, $p_{0}$ is a reference pressure 
(by convention taken to be 1 bar, as will be done here), and $\mu^{0}$ is the 
chemical potential per molecule measured at the reference pressure,
\begin{equation}
\mu^{0}(T)=h-Ts\mid_{p=1},
\label{eq:Mu0SDimer}
\end{equation}
where H and S are obtained from the NIST-JANAF tables~\cite{JANAF}. $\mu^{0}$ obtained from the JANAF tables has a physically unrealistic infinite 
value at T=0K due to the presence of 1/T terms in both the H and TS contributions.  
As shown in Figure \ref{fig:JANAFfit}, $\mu^{0}$ is nearly linear for temperatures of interest,
so we fit to a linear equation using the $770K < T < 1430K$ region to eliminate the T=0K pole.  There is still 
an issue of constant offsets to energy; the JANAF tabulated values are relative to the enthalpy at T=298K, and whereas all our calculated 
values are computed at T=0K.  In order to incorporate the empirical data into our method, we must shift its zero of energy.  
In particular, at T=0K, we set the empirical data to match what first principles calculations would have predicted,
\begin{equation}
g_{S_{2}}\mid_{T=0K}=e_{config}+e_{phonon}\mid_{T=0K}.
\label{eq:offsetGSDimer}
\end{equation}
As we have measured all energies relative to the tie-plane between pure elements, $e_{config}\equiv 0$, and so the zero-point energy $e_{phonon}$
is the only contribution at T=0K taken from first principles calculations.

Shown in Table \ref{tab:predom} are our predicted sulfidization thresholds relative to experimental predominance 
diagrams obtained by Taylor~\cite{Taylor85}.  Predictions with and without phonon free energies are shown.  
It is necessary to include phonons to obtain reasonable agreement with
experimental results, as when phonon free energies are excluded, the relative errors in pressure are as large as four orders of magnitude.
When including phonon free energies, in all cases our pressure estimates are within a factor of 6 of experimental results.  It should also be be 
noted that Taylor's phase boundaries are themselves based on fits from 
experimental enthalpy data.  One difficulty is that 
Taylor has a liquid region in the high-temperature, low pressure region of the 
predominance diagram, whereas we are only interested in solid phases,
so experimental thresholds from the solid regions were extrapolated 
to liquid regions to give numbers suitable for comparison.  This extrapolation 
is valid as Taylor's thresholds in the solid regions were found to be extremely 
linear.  Another difficulty is that Pd\subscript{3}S is stable 
for a small pressure range in between temperatures of 830K and 920K, and it would 
likely be stable for higher temperatures if not for the liquid phase, but our 
calculations do not predict it stable in this region.  This is most likely due 
to errors in the phonon free energy, which as already shown appreciably affects results.

\begin{center}
	\begin{table}
	\begin{tabular}{ |c|c|c|c|c| }
		\hline
		T (K) & Transition & $log_{10}\sqrt{P}$ Calculated & $log_{10}\sqrt{P}$ Calculated &  $log_{10}\sqrt{P}$ Experimental~\cite{Taylor85} \\
		& & Without Phonons & With Phonons & \\
		\hline
		1000 & PdS to Pd\subscript{16}S\subscript{7} & -1.1 & -1.8 & -2.2* \\
		\hline
		1000 & Pd\subscript{16}S\subscript{7} to Pd\subscript{4}S & -1.4 & -2.6 & -2.4* \\
		\hline
		1000 & Pd\subscript{4}S to Pd & -3.3 & -5.3 & -5.2 \\
		\hline
		800 & PdS to Pd\subscript{16}S\subscript{7} & -3.1 & -3.5 & -3.7 \\
		\hline
		800 & Pd\subscript{16}S\subscript{7} to Pd\subscript{4}S & -3.5 & -4.5 & -4.7 \\
		\hline
		800 & Pd\subscript{4}S to Pd & -5.9 & -7.6 & **\\
		\hline
	\end{tabular}
	\caption{Sulfidization threshold pressures for the Pd-S\subscript{2} system.  Calculated values with and without phonon contributions are separately shown.  Values * are liquid experimentally, but were extrapolated from the solid region of the predominance diagram for comparison purposes.  ** not indicated in experimental results. }
	\label{tab:predom}
	\end{table}	
\end{center}

\subsection{Activity Calculations}
In this last section, the activities of species in binary Cu-Pd are calculated, 
as these can be used to predict sulfidization thresholds for Cu-Pd in the
presence of S.  We wish here to find the activities across the entire composition 
range with multiple phases.

The definition for the activity of a species A as a function of composition $x_{A}$, pressure P, and temperature T is~\cite{ThermoDeHoff}
\[ \log a_{A}(x_{A},P,T) = \frac{\mu_{A}(x_{A},P,T)-\mu_{A}^{0}(P,T)}{k_{B}T},\]
where $a_{A}$ is the activity for species A, $\mu_{A}(x_{A})$ is the chemical potential for species A, and $\mu_{A}^{0}$ is a reference
chemical potential for species A.  As we are specifying $x_{A}$, P, and T, the natural choice of thermodynamic potential is the
Gibbs free energy,
\[G=G(N_{A},N_{B},p,T)=\mu_{A}N_{A}+\mu_{B}N_{B},\]
where $N_{A}$ and $N_{B}$ are the total number of A and B atoms, respectively, and the thermodynamic derivative giving the chemical potential is
\[\mu_{A}(N_{A},N_{B},P,T)=\frac{\partial G}{\partial N_{A}}\mid_{N_{B},P,T}.\]
However, the quantity of importance for phase transitions is the Gibbs free energy per atom,
\[g=g(x_{A},P,T)=\frac{G(N_{A},N_{B},P,T)}{N_{A}+N_{B}}.\]
We may express the chemical potential in terms of intensive quantities as
\[\mu_{A}(x_{A},P,T)=g+(1-x_{A})\frac{\partial g}{\partial x_{A}}\mid_{P,T},\]
where $x_{i}$ is the composition of species i, and we have chosen to parameterize the free energy by the composition
for A.  Similarly,
\[\mu_{B}(x_{A},P,T)=g+(1-x_{B})\frac{\partial g}{\partial x_{B}}\mid_{P,T}=g-x_{A}\frac{\partial g}{\partial x_{A}}\mid_{P,T}.\]
These activities depend only on the free energy and the derivative of the free energy at
a given composition.  They are piecewise analytic functions across the composition range,
completely determined by the thermodynamically stable phase at that composition.  The activities
can therefore be viewed as alternative representation of the underlying phase diagram for the system.  They are
discontinuous only when the derivative of the free energy changes (the free energy
itself must always be continuous, due to the requirement of convexity).

As mentioned before, we believe the magnitude of the calculated free energy of our fcc phase is artificially
large, destabilizing the $\beta$ phase, and thus any calculated activities would be
flawed near the 50\% Cu region of the phase diagram.  To support this assertion, here 
we refit the fcc phase using only low-solubility first principles data on either side of the composition range.
This reduces the magnitude of the free energy of the fcc phase while still using first principles 
derived calculations.  This should not be viewed as abandoning our pure first principles phase 
diagram as shown before, but rather a correction inspired by known experimental results presented in
an alternative-but-equivalent format.  The SQS estimate for fcc at 50\% Pd lies almost directly on this free energy
curve, further supporting its usage for quantitative estimates.  We also neglect phonon free energy 
for activity calculations, as for ordered fcc and bcc at 50\% Pd, the difference in phonon free 
energy per atom is only 2 meV at T=1000K.

Shown below in Figure \ref{fig:cupdactivities} in solid lines are our calculated activities 
for palladium in the Cu-Pd binary, using the free energy models from Table \ref{tab:phaseeqs}, 
the modified fcc phase, and our calculated data for Cu\subscript{3}Pd.  
The $\beta$ phase is the sharply increasing region in the center of the graph.
This sharp behavior is due to the entropic portion of the free energy, which has a 
logarithmic singularity at x=0.5 and sharply varying derivative near this singularity.  
The smooth curves on the left and right sides of the graph for T$<$1000K denote the fcc phase.
Linear free energy functions give constant activities as a function of composition, 
and thus phase coexistence regimes appear as plateaus in activities, at a fixed temperature.  
Two sets of plateaus are present.  The set of plateaus on 
the left side are coexistence regimes between the bcc and
fcc phases.  There are 3 types of plateaus on the right side.  For T $<$ 900K, the 
plateaus are due to coexistence with the line compound Cu\subscript{3}Pd, where the 
plateaus in the region $0.6 < x < 0.8$ indicate coexistence with bcc and the plateaus 
at $0.8 < x < 0.9 $ indicate coexistence with fcc.  The plateau on the right side of 
the plot for T=900K is a special case.  Our free energy models predict that the 
Cu\subscript{3}Pd phase becomes thermodynamically unstable at 899K and fcc becomes stable, so the
plateau at 900K indicates coexistence between bcc and fcc.  At a temperature of 960K, 
our model predicts that the bcc phase becomes unstable relative to the fcc phase, 
disappearing at a composition of x=0.576.  By decreasing the fcc free energy magnitude we have
properly stabilized the $\beta$ phase at temperatures of interest and brought the 
transition temperature to within 90K of the expected value.  The 1000K and 1350K plots are 
pure fcc throughout the entire composition range.  

The activities of Pd at various temperatures were also calculated using CALPHAD (acronym of CALculation of PHAse Diagrams) 
method. The thermodynamic database developed by Li et al\cite{Li08} was used.  Our activities are systematically low 
compared to the CALPHAD activities (shown in dashed lines on Figure \ref{fig:cupdactivities}), most likely due to our ideal 
entropy approximation.  As the true entropy in solid solutions is reduced relative to the ideal entropy, the 
true free energy will be greater, leading to the true activities being greater than our calculated activities.  
This is especially true for fcc, where the refitted free energy still overestimates the magnitude of the free 
energy and leads to qualitatively different behavior in the high Pd limit.  CALPHAD results give
activities that exceed Raoult's Law in the high Pd limit, which would require positive enthalpies 
of formation for Cu substitution in Pd using a solid substitution model in the 
dilute limit.  This contradicts first principle calculations which unambiguously predict 
Cu substitutions in bulk Pd having negative enthalpies of formation (see Figure \ref{fig:binaryphase}).  
Clearly more precise models are needed to understand the binary in the high Pd 
limit.  However, for the $\beta$ phase and associated coexistence regimes, the first principles 
and CALPHAD results are in close agreement, and if one is interested in regions outside the single-phase
fcc region, first principles calculations gives reasonably accurate results 
even with simple models. 

\begin{figure}
	\includegraphics[trim = 0mm 0mm 0mm 10mm, clip, width=\textwidth]{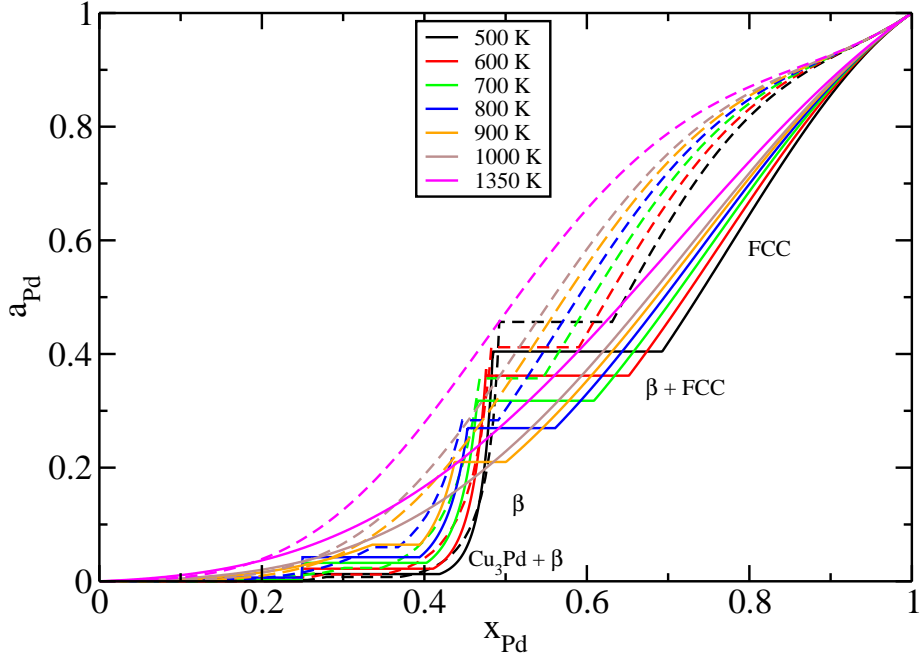}
	\caption{(color online) Activities of Pd in the Cu-Pd binary.  First principles results are denoted with solid lines, and CALPHAD with dashed lines.}
	\label{fig:cupdactivities}
\end{figure}

\section{Conclusions}
Using only first principles calculations and solid solution models, 
we were able to semi-quantitatively model the ternary phase diagram for the 
Cu-Pd-S system without any empirical fitting parameters.  We obtained the sulfidization 
thresholds for the Pd-S\subscript{2} system and activities that reflect the 
essential features of the Cu-Pd binary system at temperatures where multiple phases exist across the 
composition range.  More detailed models are necessary to reach 
better quantitative agreement with experimental results.   However, our 
work reveals the essential physics underlying the phase behavior.  For example, 
we distinguish enthalpic vs. entropic driven substitution, and we
identify a pseudogap stabilization mechanism, yielding a temperature-independent 
solubility range for Cu in Pd\subscript{16}S\subscript{7}.

\section*{Acknowledgements}
We are grateful to B. Gleeson, M. Mihalkovi\v{c}, and J. Kitchin for useful discussion.  This technical effort was 
performed in support of the Fuels Program of Strategic Center for Coal at DOE National 
Energy Technology Laboratory under the RES contract DE-FE0004000.  


\end{document}